\documentclass[3p,times,procedia]{elsarticle}
\usepackage{nupha_ecrc}
\usepackage{colortbl}
\usepackage{hyperref}
\usepackage{lineno}
\usepackage{graphicx,subfigure}
\usepackage{color}
\usepackage{pifont}
\usepackage{makecell}
\definecolor{dgreen}{cmyk}{1.,0.,1.,0.2}        
\definecolor{orange}{cmyk}{0.,0.353,1.,0.}    
\usepackage{xcolor,colortbl}
\usepackage{diagbox}
\usepackage[utf8]{inputenc}

\definecolor{Gray}{gray}{0.85}
\definecolor{LightCyan}{rgb}{0.88,1,1}
\definecolor{LightBlue}{rgb}{0.7,0.9,1}
\definecolor{LightGreen}{rgb}{1,0.9,1}

\newcolumntype{a}{>{\columncolor{Gray}}c}
\newcolumntype{b}{>{\columncolor{white}}c}

\volume{00}

\firstpage{1}

\journalname{Nuclear Physics A}

\runauth{You Zhou}

\jid{nupha}

\jnltitlelogo{Nuclear Physics A}

\usepackage{amssymb}

\biboptions{sort&compress}

\usepackage[figuresright]{rotating}

\begin{document}

\begin{frontmatter}




\dochead{XXVIIth International Conference on Ultrarelativistic Nucleus-Nucleus Collisions\\ (Quark Matter 2018)}

\title{Collective Effects in Nuclear Collisions: Experimental Overview}
\author{You Zhou}
\ead{you.zhou@cern.ch}
\address{Niels Bohr Institute, University of Copenhagen, Blegdamsvej 17, 2100 Copenhag1en, Denmark}


\author{}

\address{}

\begin{abstract}
These conferences proceedings summarize the experimental findings of collective effects in nuclear collisions, as presented at the Quark Matter 2018 conference.
\end{abstract}


\end{frontmatter}


\section{Introduction}
\label{sec:intro}

One of the main goals of the ultra-relativistic heavy-ion collision programme at the Relativistic Heavy Ion Collider (RHIC) and the Large Hadron Collider (LHC) is to create a novel state of strongly interacting matter named the Quark-Gluon Plasma (QGP), and to study its properties of the novel state of strongly interacting matter that is proposed to exist at high temperatures and energy densities~\cite{Shuryak:1978ij}. Studies of azimuthal correlations of produced particles have contributed significantly to the characterisation of the matter created in heavy-ion collisions~\cite{Ollitrault:1992bk,Voloshin:2008dg}. Anisotropic flow, which quantifies the degree of anisotropic azimuthal correlations of final state particles, is found to be very sensitive both to the event-by-event fluctuating initial geometry of the overlap region of colliding nucleons and to the transport properties and its equation of state~\cite{Voloshin:2008dg, Heinz:2013th,Song:2017wtw}. The successful descriptions of anisotropic flow measurements by hydrodynamic model calculations suggest that the created hot and dense medium behaves like a nearly-perfect fluid~\cite{Voloshin:2008dg,Heinz:2013th} with a shear viscosity over entropy density ratio, $\eta/s$, close to the conjectured lower bound of $1/4\pi$~\cite{Kovtun:2004de}. Anisotropic flow can be characterised using a Fourier decomposition of the azimuthal distribution of the emitted particles in the plane transverse to the beam axis~\cite{Voloshin:1994mz, Poskanzer:1998yz}: 
\begin{equation}
\frac{dN}{d\varphi} \propto 1+2\sum_{n=1}^{\infty}v_{n} \cos[n(\varphi - \Psi_{n})],
\label{eq:FourierExp}
\end{equation}  
where $N$ is the number of produced particles, $\varphi$ is the azimuthal angle of each particle and $\Psi_{n}$ is the $n^{th}$ order flow symmetry plane. The $n^{th}$ order (complex) anisotropic flow $V_{n}$ is defined as: $V_{n} \equiv v_{n} \, e^{in\Psi_{n}}$, and $v_{n}  = |V_{n}|$ is the flow coefficient which gives the magnitude of the flow-vector, and $\Psi_{n}$ represents the orientation of the flow-vector. For non-central heavy-ion collisions, the dominant flow coefficient $v_2$ referred to as elliptic flow, has been investigated for several decades. Investigations of higher order coefficients (especially odd orders), developed rapidly after 2010. The observed non-zero values of higher flow coefficients $v_n$ at RHIC and the LHC are ascribed primarily to the response of the produced QGP to fluctuations of the initial geometry fluctuations~\cite{Alver:2010gr, ALICE:2011ab, ATLAS:2012at, Chatrchyan:2013kba, Adam:2016izf}. 

In addition to standard studies on flow coefficient for charged and identified particles, other flow observables, for example event-by-event flow fluctuations~\cite{Aad:2013xma,Gale:2012rq}, fluctuations (de-correlation) of flow-vector~\cite{Aad:2014fla,Qiu:2012uy, Heinz:2013bua,Gardim:2012im,Khachatryan:2015oea,Acharya:2017ino}, and the correlations between different order flow-vectors~\cite{Aad:2015lwa,ALICE:2016kpq,Giacalone:2016afq,Zhu:2016puf,Qian:2016pau,Zhou:2015eya}, have also been investigated in great details in relativistic heavy-ion collisions. 
The comparisons between latest experimental data and recent development of event-by-event hydrodynamical model simulations on various flow observables shed new insight into the the initial conditions and the properties of the create QGP matter in heavy-ion collisions~\cite{Song:2017wtw}.


In this proceedings contribution, I will give an experimental overview of the latest anisotropic flow measurements in nuclear collisions, with a preferred selection from new analyses presented at Quark Matter 2018.

\section{Anisotropic flow coefficient}
\label{}

Nearly-all experiments performed the measurements of anisotropic flow coefficients $v_n$ at RHIC and the LHC. Early measurements of $v_2$ in Au--Au collisions at $\sqrt{s_{_{\rm NN}}} = $ 130 GeV and comparisons to ideal hydrodynamic calculations showed some of the first evidence of the creation of a QGP and suggested that this matter behaved nearly as a perfect fluid. Later measurements, especially with an improved understanding of event-by-event geometric fluctuations, significantly advanced the extraction of QGP transport coefficients and provided constraints on the initial conditions. The new data from LHC Xe--Xe run completed in 2017 provide a new opportunity to constrain $\eta/s$ further. The first anisotropic flow measurements with 2- and multi-particle cumulant methods in Xe--Xe collisions at $\sqrt{s_{_{\rm NN}}} = $ 5.02 GeV are reported by ALICE in~\cite{Acharya:2018ihu}. A stronger centrality dependence of $v_{2}$ is observed, while the dependence is weaker for $v_3$ and $v_4$. This is expected as the change of $\varepsilon_2$ from central to peripheral collisions is more dramatic than $\varepsilon_3$ and $\varepsilon_{4}$ (for $v_{4}$, $\varepsilon_{4}$ is the main contribution for non-peripheral collisions). Similar results were observed in Pb--Pb collisions as well~\cite{Acharya:2018lmh}. Besides, the difference of $v_2$ from 2- and multi-particle cumulants has been seen, which is mainly because of the event-by-event fluctuations of $v_{2}$. However, the possible difference between $v_{2}$ from higher order cumulants, e.g. $v_{2}\{4\}$, $v_{2}\{6\}$ and $v_{2}\{8\}$, is helpful in the investigations on underlying probability density distribution of $v_{n}$. However, it is difficult to measure such a difference due to the limited statistics. ALICE also presents the comparisons of $v_{n}$ to hydrodynamic predictions using EKRT initial conditions and two parameterizations of $\eta/s$(T). It is observed that the hydrodynamic predictions describe the data fairly well from central to mid-peripheral collisions. This agreement between experimental data and theoretical predictions, on the one hand, confirms that the created strongly interacting QGP matter in Xe--Xe collision behaves like a perfect fluid. On the other hand, it also shows that with $v_{n}$ alone one is not able to discriminate two different functional forms of $\eta/s(T)$. It is obvious that we need tighter constraints on QGP properties than just the centrality dependence of $v_n$. One step further has been done by looking at the differential measurements of $v_n$. The $p_{\rm T}$-differential measurement have been presented by the ALICE, ATLAS and CMS collaborations~\cite{Acharya:2018ihu,ATLAS-CONF-2018-011,CMS-PAS-HIN-18-001}. For example, the ATLAS collaboration has shown measurements of $v_{n}(p_{\rm T})$ for $n=$ 2--7~\cite{ATLAS-CONF-2018-011}. A clear ordering of $v_2 > v_3 >  ... > v_7$ is observed. This characteristic flow feature is also confirmed by other collaborations~\cite{Acharya:2018ihu,CMS-PAS-HIN-18-001}. At the time of Quark Matter 2018 conference, however, no direct data and theory comparison are available for $v_{n}(p_{\rm T})$ in Xe--Xe collisions. It is without any doubt that future developments in theoretical calculations, especially the Global Bayesian Analysis with combined Pb--Pb and Xe--Xe flow data with a common parameterization of transport coefficients, will significantly advance our overall understanding of QGP created in relativistic heavy-ion collisions. 

Last but not least, the 8-hours Xe--Xe data taking at the LHC provides many new possibilities of system size dependence of flow observables. Only very first studies of $v_{n}$ coefficient are available at the moment while prospective analysis on other observables will undoubtedly tighten the constraints on initial conditions of heavy-ion collisions and shed more insight into the nature of QGP creation and evolution. This should also trigger the considerations of future possibilities of different colliding nuclei provided at the LHC. One example could be Oxygen, which has been discussed this year during and after the Quark Matter 2018 conference, but which will have to wait until LHC Run 3.

\section{Flow fluctuations}
\label{}

The extraction of precise information on QGP properties requires a good knowledge of the initial conditions of heavy-ion collisions, which contribute to the dominant uncertainty of the extracted transport coefficients. One of the previous examples is the extracted $\eta/s$ from $v_{2}$ measurement, it was 0.08 from hydrodynamic calculations with MC-Glauber initial conditions, and was 0.16 using MC-KLN initial conditions, which gives a 100\% uncertainty of the extracted $\eta/s$. The initial event-by-event geometry fluctuations not only results in fluctuations of even order flow coefficients but also leads to non-zero odd order flow coefficients. This initial fluctuations lead to $\langle v_{n}^{k} \rangle \neq \langle v_{n}\rangle^{k}$ and the development of different order flow symmetry planes $\Psi_{n}$ (which typically, but do not necessarily coincide with the reaction plane defined by the impact parameter and beam direction). A comprehensive understanding of both initial state geometry fluctuations and their effects on the final state anisotropic flow is crucial for the comparison of flow measurements to the theoretical descriptions. In this proceedings, two main questions will be discussed in detail, namely: \\
$~~~~~~  \bullet$ How do $\varepsilon_{n}$ and $v_{n}$ fluctuate and what is the underlying probability density function (PDF) of their distributions; \\
$~~~~~  \bullet$ How are the initial geometry fluctuations reflected in differential flow measurements. Is there de-correlation of flow-vector in $p_{\rm T}$ or $\eta$ intervals.

Several possible candidates for the underlying PDF of $v_{n}$ or $\varepsilon_{n}$ are proposed. A commonly used parameterisation of $P(\varepsilon_{n})$ is the Bessel-Gaussian distributions~\cite{Voloshin:2007pc}. 
Based on initial state model studies, it was found that the Bessel-Gaussian function describes the $\varepsilon_{2}$ distribution well in mid-central collisions. However, it does not fit the distribution in peripheral collisions, due to it lacking a constraint of $\varepsilon_{2} <$ 1~\cite{Yan:2014afa}. 
In 2014, the Elliptic-Power function was proposed as the description of underlying PDF~\cite{Yan:2014afa}. This function not only mimics the initial density profile from MC-Glauber model but also quantitatively fits the $p(\varepsilon_{n})$ distributions from various initial state models~\cite{Yan:2014afa,Zhou:2015eya}. It shows that the Elliptic-Power function could be a better candidate than the Bessel-Gaussian function to model the PDF of $\varepsilon_n$.

While discussing event-by-event $v_n$ distributions, flow measurements for charged and identified particles were initially used to study the mean value of the PDF, which is the first moment of the $P(v_{n})$ distribution. Later on, with $v_2$ from 2- and 4-particle cumulants, the second moment was studied via the relative fluctuations $\sigma_{v_n}/ \left< v_n \right> = \sqrt{\frac{v_{n}\{2\}^{2} + v_{n}\{4\}^{2}}{v_{n}\{2\}^{2} - v_{n}\{4\}^{2}} }$. 
However, with only the first and second moment, it is difficult to describe a broad range of dynamic phenomena since different PDFs could accidentally give the same lower order moments. Full investigations on the underlying PDF are therefore necessary, and there are many recent developments.
Two different approaches have been used by the experiments. One is using the Bayesian unfolding technique to reconstruct the PDF directly, while the other is trying to reconstruct the underlying PDF by as many moments, or equivalently cumulants, as possible.
The first one was done by ATLAS collaboration in Pb--Pb collisions at 2.76 TeV (LHC Run 1)~\cite{Aad:2013xma}, and this was followed up in a recent work by CMS using high statistics LHC Run 2 data~\cite{Sirunyan:2017fts}. In both measurements, a non-Gaussian behaviour is observed in the event-by-event fluctuations of $v_2$. The underlying PDF of $v_2$ distribution could be fitted well by both Bessel-Gaussian and Elliptic Power function for central collisions, while for non-central collisions the data prefer descriptions using the Elliptic Power function. In the second approach, in addition to the study of first and second moments, the third order moment (skewness), was obtained from 4- and 6-particle cumulants. It is found to be negative with an increasing magnitude as collisions become less central. Using the relation between multi-particle cumulants of $v_2$ and the Elliptic Power parameters given in~\cite{Yan:2014afa}, the underlying PDF of $v_2$ is constructed. The resulting $P(v_2)$ scaled by its mean value $\left< v_2 \right>$ is in a good agreement with the results reported by ATLAS using the first approach, despite the different kinematic cuts and collision energies used the by two experiments.

Together with many other studies, we conclude the following:\\
$~~~~~~  \bullet$ Bessel-Gaussian: predicts $v_{3}\{m\} = v_{3}\{\Psi_{RP}\} =$ 0 and $\varepsilon_{3}\{m\} = \varepsilon_{3}\{\Psi_{RP}\} = 0$ (for $m \geq$ 4), which are not supported by the experimental data and model calculations. In addition this function can not fit $P(\varepsilon_{n})$ and $P(v_n)$ distributions in non-central collisions; \\
$~~~~~~  \bullet$ Elliptic-Power: quantitatively reproduces both the $P(\varepsilon_{n})$ and $P(v_n)$ distributions in all centrality intervals, predicts non-zero value of  $v_{3}\{m\}$ and $\varepsilon_{3}\{n\}$ ($m \geq$ 4), and predicts $v_{n}\{4\} > v_{n}\{6\} > v_{n}\{8\}$, which has been confirmed in data. \\
From the above, it is clearly that Elliptic-Power function is a promising candidate of the underlying PDF of $v_{n}$ distributions. 

The study of flow fluctuations provides insight into geometry fluctuations and their effects on the final anisotropic flow measurements. However, the comparison of anisotropic flow measurements and hydrodynamic calculations are not only applied for the integrated, but also for the differential flow, which is more sensitive to $\eta/s$ and the initial conditions.
Additional $p_{\rm T}$ and $\eta$ dependent fluctuations of symmetry plane orientation (i.e. the flow angle) and magnitude might be produced by the response of the system to initial geometry fluctuations~\cite{Heinz:2013bua, Gardim:2012im}. The systematic studies of $p_{\rm T}$ and $\eta$ fluctuations have been performed in great detail at the LHC, via $v_{n}\{2\}/v_{n}[2]$ which mixes reference and differential correlated particles from various kinematic ranges~\cite{Acharya:2017ino} and the factorization ratio $r_n$ in Refs.~\cite{Khachatryan:2015oea,Acharya:2017ino,Aaboud:2017tql}. It is reported that the 2-particle function measured in two separated kinematic intervals ($p_{\rm T}$ or $\eta$) does not factorize into the product of $v_n$ from each kinematic interval. This is often called single-particle $v_n$, a name which is insufficiently precise because $v_n$ quantifies the anisotropic azimuthal correlations among many particles. The flow-vector de-correlation (factorization breakdown) becomes larger when increasing the $p_{\rm T}$ or $\eta$ separation between the kinematic intervals. A nice attempt has been made by the ATLAS Collaboration to separate the contributions from forward-backwards asymmetry with the event-plane twist in de-correlation effects~\cite{Aaboud:2017tql}. It was found that the two contributions are comparable. Furthermore, the de-correlation effect is found to be larger at 2.76 TeV than 5.02 TeV, which can not be simply explained by the change in the beam rapidity. An important measurement was done recently at RHIC by the STAR Collaboration~\cite{Maowu:2018QM}. They further confirm the collision energy dependence of the de-correlation effect, with an even stronger effect observed in 200 GeV Au--Au collisions. This measurement, unfortunately, can not be described quantitatively by the recently developed (3$+$1)D CLVisc hydrodynamic model~\cite{Wu:2018cpc}.

\section{Flow Correlations}
\label{}

The correlation between different order flow coefficients is one of the two key observables related to flow-vector correlations, besides the correlations between flow angles. It was first measured by the ATLAS Collaboration using Event-Shape Engineering (ESE)~\cite{Schukraft:2012ah}. It was found that there is an anti-correlation between $v_{2}$ and $v_{3}$, and a positive correlation between $v_{2}$ and $v_{4}$~\cite{Aad:2015lwa}. To quantify the correlation strength, a novel observable called Symmetric Cumulants $SC(m, n)$ was proposed, and it could be measured by the multi-particle cumulant technique~\cite{Bilandzic:2013kga}. By design, this observable should be independent on the symmetry plane correlations, and it is expected to be insensitive to non-flow effects, which was confirmed by a Monte-Carlo study using HIJING simulations~\cite{ALICE:2016kpq}.

The first experimental measurements of $SC(m, n)$ was done by ALICE, where positive values of $SC(4, 2)$ and negative values of $SC(3, 2)$ were reported, consistent with the early ESE studies by ATLAS~\cite{Aad:2015lwa}. Besides, it was found that although hydrodynamic calculations using EKRT initial condition and various parameterizations could nicely describe measurements of $v_n$, none of the hydrodynamic calculations could quantitatively fit the sizes of $SC(3,2)$ and $SC(4,2)$ simultaneously. This study was further extended to correlations between higher order flow coefficients by ALICE~\cite{Acharya:2017gsw} and compared to more available theoretical calculations~\cite{Zhu:2016puf}, which again suggest $SC(m, n)$ as a tight constraint on theoretical models. Recently, results are also reported in various collision systems, including Pb--Pb, Au--Au, Xe--Xe, p--Pb and pp collisions~\cite{Sirunyan:2017uyl,Gajdosova:2018pqo}.

Studies of  non-linear hydrodynamic response were made already many years ago in both theoretical calculations and experimental measurements (e.g. by looking at $v_{4}$ from the second-order event-plane ${\rm EP_{2}}$, $v_{4}\{\rm EP_{2}\}$, at RHIC~\cite{Adams:2003zg}). Later on, the ATLAS Collaboration extracted the linear and non-linear components using Event-Shape Engineering~\cite{Aad:2015lwa} and also performed an investigation of event-plane correlations~\cite{Aad:2014fla}. These efforts were followed by the ALICE Collaboration using a modern multi-particle correlation with {\it Generic Framework} method in Ref.~\cite{Acharya:2017zfg}. It was found that higher order anisotropic flow can be isolated into two independent contributions: a component that arises from a non-linear response of the system to the lower order initial anisotropy coefficients $\varepsilon_{2}$ and/or $\varepsilon_{3}$, and a linear mode which is driven by linear response of the system to the same order cumulant-defined anisotropy coefficient. The non-linear flow mode could nicely explain the correlations between different order flow symmetry planes $\rho_{n,mk}$ (or named ``Event-Plane correlation''). The first investigation on non-linear flow mode coefficient $\chi_{n,mk}$ showed that this type of observables has different sensitivity to $\eta/s$ and the initial conditions. For example, $\chi_{4,22}$ has a strong dependence on the initial conditions, but it is almost insensitive to $\eta/s$. This suggests that $\chi_{4,22}$ measurement can be used to constrain the initial conditions, with less sensitivity to the values of $\eta/s(T)$ in hydrodynamic calculations than other flow observables. On the other hand, $\chi_{5,32}$ and $\chi_{6,33}$ are only weakly sensitive to the initial conditions but depend strongly on $\eta/s$ at the freeze-out condition. The published ALICE data~\cite{Acharya:2017zfg}, are not able to further constrain the $\eta/s$ during system evolution. However, they provide unique information on freeze-out condition which was poorly known with other anisotropic flow related observables. In this conference, ALICE moves one step further by studying the non-linear flow mode for identified particles~\cite{Naghmeh:2018QM}, including $\pi^{\pm}$, $K^{\pm}$ and $p(\bar{p})$. Interestingly, it is reported that there is a clear mass ordering of non-linear flow mode $v_{n,mk}(p_{\rm T})$ observed at low $p_{\rm T}$, which can be described by the viscous hydrodynamic calculation iEBE-VISHNU. This mass ordering at low $p_{\rm T}$ is followed by a baryon-meson grouping phenomenon at higher $p_{\rm T}$ range, often taken as evidence of partonic collectivity in relativistic heavy-ion collisions. The characteristic flow feature has been reported in many of the previous measurement of high order $v_{n}$ for identified particles, and it seems similar also for the sub-component of the total flow $v_n$. 

However, at the moment, it is not obvious how to connect the presented $v_{n,mk}(p_{\rm T})$ to differential symmetry plane correlation and non-linear flow mode coefficient, and it is unclear whether $v_{n}\{2\}$ or $v_{n}[2]$ (for $n \geq 4$) represents the flow from the corresponding flow symmetry plane of this differential study. Further studies will certainly improve the understanding of non-linear hydrodynamic response to the initial geometry and help to use the available experimental data to tighten constraints on theory. An intense discussion between experimental and theoretical physicists is ongoing.

\section{Instead of a summary}

\begin{table}[bt]
\label{my-label}
\begin{tabular}{|c|c|c|c|c|c|c|c|}
\hline
\rowcolor{Gray}
\diagbox[width=5.em]{\rule{0mm}{1.cm}\rule{0.cm}{0cm} Setting}{Model}   &   \makecell{\small iEBE-VISHNU \\ (I) \\ \small Ref.~\cite{Zhao:2017yhj} }    &  \makecell{\small iEBE-VISHNU \\ (II) \\ \small Ref.~\cite{Zhao:2017yhj}  }  &     \makecell{VISH2+1  \\ \small Ref.~\cite{Zhu:2016puf} } &  \makecell{\small EKRT \\ +Hydro \\ (fixed $\eta/s$) \\ \small Ref.~\cite{Niemi:2015voa}} &  \makecell{\small  EKRT \\ +Hydro \\ (param I) \\ \small Ref.~\cite{Niemi:2015voa} } & \makecell{\small IP-Glasma \\ + MUSIC  \\ + UrQMD \\ \small Ref.~\cite{McDonald:2016vlt}} \\ \hline 
\rowcolor{LightBlue}
Initial   conditions &  T$_{\rm R}$ENTo & AMPT & AMPT & EKRT & EKRT & IP-Glasma   \\ \hline
\rowcolor{LightBlue}
$\eta/s$ &  $\eta/s$(T) &    {\small $ \eta/s =0.20$}  &   {\small $ \eta/s =0.16$} &   {\small $ \eta/s =0.20$}  & $\eta/s$(T)  &  {\small $\eta/s= 0.095$} \\ \hline
\rowcolor{LightBlue}
$\zeta/s$ &  $\zeta/s$(T)  &  $\zeta/s$ = 0  &  $\zeta/s$ = 0  & $\zeta/s$(T)  & $\zeta/s$(T)  &   $\zeta/s$(T)  \\ \hline

\rowcolor{Gray}
Observables  \\ 
\hline 
$v_{2}$  & $\checkmark$ &  $\checkmark$ & $\checkmark$ &$\checkmark$ &$\checkmark$ &$\checkmark$  \\ \hline 
$v_{3-7}$   & $\checkmark$ &  $\checkmark$ & $\Delta$ &$\checkmark$ &$\checkmark$ &$\checkmark$  \\ \hline 
$P(v_{n})$   & $\checkmark$ &  $\checkmark$ & $\Delta$ &$\checkmark$ &$\checkmark$ &$\checkmark$  \\ \hline 
$v_{n}(p_{\rm T})^{ch, PID}$   & $\Delta$ &  $\checkmark$ & $N/A$ &$N/A$ &$N/A$ &$\Delta$  \\ \hline 
$r_{n}$   & $\Delta$ &  $\Delta$ & $N/A$ &$N/A$ &$N/A$ &$\Delta$  \\ \hline 
$SC(m,n)$   & $\Delta$ &  $\Delta$ & $\times$ &$\Delta$ &$\Delta$ &$N/A$  \\ \hline 
$v_{n,mk}$   & $\checkmark$ &  $\checkmark$ & $N/A$ &$\checkmark$ &$\checkmark$ &$\checkmark$  \\ \hline 
$\rho_{n,mk}$   & $\checkmark$ &  $\checkmark$ & $N/A$ &$\checkmark$ &$\checkmark$ &$\checkmark$  \\ \hline 
$\chi_{n,mk}$   & $\checkmark$ &  $\checkmark$ & $N/A$ &$N/A$ &$N/A$ &$\checkmark$  \\ \hline 
$v_{n,mk}(p_{\rm T})^{ch, PID}$   & $\Delta$ &  $\checkmark$ & $N/A$ &$N/A$ &$N/A$ &$N/A$  \\ \hline 
\end{tabular}
\small
\caption{Current available comparisons of between data and model calculations. Here $\checkmark$  (Good), $\Delta$ (Not so bad), $\times$ (Not good) and $N/A$ (Not available).}
\end{table}

Instead of a summary, the current comparisons of experimental measurements on anisotropic flow related observables and theoretical calculations are presented in Table~\ref{my-label}. Note that only a small selection of model calculations are listed here, and the table represents personal opinions only. It is clear that hydrodynamic models with specially tuned parametrizations can qualitatively or even quantitatively describe a large set of experimental data. Nevertheless, there is still room to further improve model calculations, especially with Global Bayesian Analysis (after combining Pb--Pb and Xe--Xe fits or calibrating model parameters with more differential measurements). Last but not least, it has been noticed in the Quark Matter 2018 conference that Deep Learning has become a topic of great interest in the heavy-ion community. The results presented in Refs.~\cite{Pang:2016vdc,Huang:2018fzn} start to show that Deep Learning might have the unique potential to be a powerful tool to probe the Quark-Gluon Plasma.

\section*{Acknowledgments}

The author would like to thank the organizers of Quark Matter 2018 for such an excellent conference and the invitation to give an overview talk. This work is supported by the Danish Council for Independent Research, Natural Sciences, the Danish National Research Foundation (Danmarks Grundforskningsfond) and the Carlsberg Foundation (Carlsbergfondet).







\begin{thebibliography}{10}
\expandafter\ifx\csname url\endcsname\relax
  \def\url#1{\texttt{#1}}\fi
\expandafter\ifx\csname urlprefix\endcsname\relax\def\urlprefix{URL }\fi
\expandafter\ifx\csname href\endcsname\relax
  \def\href#1#2{#2} \def\path#1{#1}\fi

\bibitem{Shuryak:1978ij}
E.~V. Shuryak, {Quark-Gluon Plasma and Hadronic Production of Leptons, Photons
  and Psions}, Phys. Lett. 78B (1978) 150, [Yad. Fiz.28,796(1978)].
\newblock \href {http://dx.doi.org/10.1016/0370-2693(78)90370-2}
  {\path{doi:10.1016/0370-2693(78)90370-2}}.

\bibitem{Ollitrault:1992bk}
J.-Y. Ollitrault, {Anisotropy as a signature of transverse collective flow},
  Phys. Rev. D46 (1992) 229--245.
\newblock \href {http://dx.doi.org/10.1103/PhysRevD.46.229}
  {\path{doi:10.1103/PhysRevD.46.229}}.

\bibitem{Voloshin:2008dg}
S.~A. Voloshin, A.~M. Poskanzer, R.~Snellings, {Collective phenomena in
  non-central nuclear collisions}, Landolt-Bornstein 23 (2010) 293--333.
\newblock \href {http://arxiv.org/abs/0809.2949} {\path{arXiv:0809.2949}},
  \href {http://dx.doi.org/10.1007/978-3-642-01539-7_10}
  {\path{doi:10.1007/978-3-642-01539-7_10}}.

\bibitem{Heinz:2013th}
U.~Heinz, R.~Snellings, {Collective flow and viscosity in relativistic
  heavy-ion collisions}, Ann. Rev. Nucl. Part. Sci. 63 (2013) 123--151.
\newblock \href {http://arxiv.org/abs/1301.2826} {\path{arXiv:1301.2826}},
  \href {http://dx.doi.org/10.1146/annurev-nucl-102212-170540}
  {\path{doi:10.1146/annurev-nucl-102212-170540}}.

\bibitem{Song:2017wtw}
H.~Song, Y.~Zhou, K.~Gajdosova, {Collective flow and hydrodynamics in large and
  small systems at the LHC}, Nucl. Sci. Tech. 28~(7) (2017) 99.
\newblock \href {http://arxiv.org/abs/1703.00670} {\path{arXiv:1703.00670}},
  \href {http://dx.doi.org/10.1007/s41365-017-0245-4}
  {\path{doi:10.1007/s41365-017-0245-4}}.

\bibitem{Kovtun:2004de}
P.~Kovtun, D.~T. Son, A.~O. Starinets, {Viscosity in strongly interacting
  quantum field theories from black hole physics}, Phys. Rev. Lett. 94 (2005)
  111601.
\newblock \href {http://arxiv.org/abs/hep-th/0405231}
  {\path{arXiv:hep-th/0405231}}, \href
  {http://dx.doi.org/10.1103/PhysRevLett.94.111601}
  {\path{doi:10.1103/PhysRevLett.94.111601}}.

\bibitem{Voloshin:1994mz}
S.~Voloshin, Y.~Zhang, {Flow study in relativistic nuclear collisions by
  Fourier expansion of Azimuthal particle distributions}, Z. Phys. C70 (1996)
  665--672.
\newblock \href {http://arxiv.org/abs/hep-ph/9407282}
  {\path{arXiv:hep-ph/9407282}}, \href
  {http://dx.doi.org/10.1007/s002880050141} {\path{doi:10.1007/s002880050141}}.

\bibitem{Poskanzer:1998yz}
A.~M. Poskanzer, S.~A. Voloshin, {Methods for analyzing anisotropic flow in
  relativistic nuclear collisions}, Phys. Rev. C58 (1998) 1671--1678.
\newblock \href {http://arxiv.org/abs/nucl-ex/9805001}
  {\path{arXiv:nucl-ex/9805001}}, \href
  {http://dx.doi.org/10.1103/PhysRevC.58.1671}
  {\path{doi:10.1103/PhysRevC.58.1671}}.

\bibitem{Alver:2010gr}
B.~Alver, G.~Roland, {Collision geometry fluctuations and triangular flow in
  heavy-ion collisions}, Phys. Rev. C81 (2010) 054905, [Erratum: Phys.
  Rev.C82,039903(2010)].
\newblock \href {http://arxiv.org/abs/1003.0194} {\path{arXiv:1003.0194}},
  \href {http://dx.doi.org/10.1103/PhysRevC.82.039903,
  10.1103/PhysRevC.81.054905} {\path{doi:10.1103/PhysRevC.82.039903,
  10.1103/PhysRevC.81.054905}}.

\bibitem{ALICE:2011ab}
K.~Aamodt, et~al., {Higher harmonic anisotropic flow measurements of charged
  particles in Pb-Pb collisions at $\sqrt{s_{NN}}$=2.76 TeV}, Phys. Rev. Lett.
  107 (2011) 032301.
\newblock \href {http://arxiv.org/abs/1105.3865} {\path{arXiv:1105.3865}},
  \href {http://dx.doi.org/10.1103/PhysRevLett.107.032301}
  {\path{doi:10.1103/PhysRevLett.107.032301}}.

\bibitem{ATLAS:2012at}
G.~Aad, et~al., {Measurement of the azimuthal anisotropy for charged particle
  production in $\sqrt{s_{NN}}=2.76$ TeV lead-lead collisions with the ATLAS
  detector}, Phys. Rev. C86 (2012) 014907.
\newblock \href {http://arxiv.org/abs/1203.3087} {\path{arXiv:1203.3087}},
  \href {http://dx.doi.org/10.1103/PhysRevC.86.014907}
  {\path{doi:10.1103/PhysRevC.86.014907}}.

\bibitem{Chatrchyan:2013kba}
S.~Chatrchyan, et~al., {Measurement of higher-order harmonic azimuthal
  anisotropy in PbPb collisions at $\sqrt{s_{NN}}$ = 2.76 TeV}, Phys. Rev.
  C89~(4) (2014) 044906.
\newblock \href {http://arxiv.org/abs/1310.8651} {\path{arXiv:1310.8651}},
  \href {http://dx.doi.org/10.1103/PhysRevC.89.044906}
  {\path{doi:10.1103/PhysRevC.89.044906}}.

\bibitem{Adam:2016izf}
J.~Adam, et~al., {Anisotropic flow of charged particles in Pb-Pb collisions at
  $\sqrt{s_{\rm NN}}=5.02$ TeV}, Phys. Rev. Lett. 116~(13) (2016) 132302.
\newblock \href {http://arxiv.org/abs/1602.01119} {\path{arXiv:1602.01119}},
  \href {http://dx.doi.org/10.1103/PhysRevLett.116.132302}
  {\path{doi:10.1103/PhysRevLett.116.132302}}.

\bibitem{Aad:2013xma}
G.~Aad, et~al., {Measurement of the distributions of event-by-event flow
  harmonics in lead-lead collisions at = 2.76 TeV with the ATLAS detector at
  the LHC}, JHEP 11 (2013) 183.
\newblock \href {http://arxiv.org/abs/1305.2942} {\path{arXiv:1305.2942}},
  \href {http://dx.doi.org/10.1007/JHEP11(2013)183}
  {\path{doi:10.1007/JHEP11(2013)183}}.

\bibitem{Gale:2012rq}
C.~Gale, S.~Jeon, B.~Schenke, P.~Tribedy, R.~Venugopalan, {Event-by-event
  anisotropic flow in heavy-ion collisions from combined Yang-Mills and viscous
  fluid dynamics}, Phys. Rev. Lett. 110~(1) (2013) 012302.
\newblock \href {http://arxiv.org/abs/1209.6330} {\path{arXiv:1209.6330}},
  \href {http://dx.doi.org/10.1103/PhysRevLett.110.012302}
  {\path{doi:10.1103/PhysRevLett.110.012302}}.

\bibitem{Aad:2014fla}
G.~Aad, et~al., {Measurement of event-plane correlations in
  $\sqrt{s_{NN}}=2.76$ TeV lead-lead collisions with the ATLAS detector}, Phys.
  Rev. C90~(2) (2014) 024905.
\newblock \href {http://arxiv.org/abs/1403.0489} {\path{arXiv:1403.0489}},
  \href {http://dx.doi.org/10.1103/PhysRevC.90.024905}
  {\path{doi:10.1103/PhysRevC.90.024905}}.

\bibitem{Qiu:2012uy}
Z.~Qiu, U.~Heinz, {Hydrodynamic event-plane correlations in Pb+Pb collisions at
  $\sqrt{s}=2.76$ATeV}, Phys. Lett. B717 (2012) 261--265.
\newblock \href {http://arxiv.org/abs/1208.1200} {\path{arXiv:1208.1200}},
  \href {http://dx.doi.org/10.1016/j.physletb.2012.09.030}
  {\path{doi:10.1016/j.physletb.2012.09.030}}.

\bibitem{Heinz:2013bua}
U.~Heinz, Z.~Qiu, C.~Shen, {Fluctuating flow angles and anisotropic flow
  measurements}, Phys. Rev. C87~(3) (2013) 034913.
\newblock \href {http://arxiv.org/abs/1302.3535} {\path{arXiv:1302.3535}},
  \href {http://dx.doi.org/10.1103/PhysRevC.87.034913}
  {\path{doi:10.1103/PhysRevC.87.034913}}.

\bibitem{Gardim:2012im}
F.~G. Gardim, F.~Grassi, M.~Luzum, J.-Y. Ollitrault, {Breaking of factorization
  of two-particle correlations in hydrodynamics}, Phys. Rev. C87~(3) (2013)
  031901.
\newblock \href {http://arxiv.org/abs/1211.0989} {\path{arXiv:1211.0989}},
  \href {http://dx.doi.org/10.1103/PhysRevC.87.031901}
  {\path{doi:10.1103/PhysRevC.87.031901}}.

\bibitem{Khachatryan:2015oea}
V.~Khachatryan, et~al., {Evidence for transverse momentum and pseudorapidity
  dependent event plane fluctuations in PbPb and pPb collisions}, Phys. Rev.
  C92~(3) (2015) 034911.
\newblock \href {http://arxiv.org/abs/1503.01692} {\path{arXiv:1503.01692}},
  \href {http://dx.doi.org/10.1103/PhysRevC.92.034911}
  {\path{doi:10.1103/PhysRevC.92.034911}}.

\bibitem{Acharya:2017ino}
S.~Acharya, et~al., {Searches for transverse momentum dependent flow vector
  fluctuations in Pb-Pb and p-Pb collisions at the LHC}, JHEP 09 (2017) 032.
\newblock \href {http://arxiv.org/abs/1707.05690} {\path{arXiv:1707.05690}},
  \href {http://dx.doi.org/10.1007/JHEP09(2017)032}
  {\path{doi:10.1007/JHEP09(2017)032}}.

\bibitem{Aad:2015lwa}
G.~Aad, et~al., {Measurement of the correlation between flow harmonics of
  different order in lead-lead collisions at $\sqrt{s_{NN}}$=2.76 TeV with the
  ATLAS detector}, Phys. Rev. C92~(3) (2015) 034903.
\newblock \href {http://arxiv.org/abs/1504.01289} {\path{arXiv:1504.01289}},
  \href {http://dx.doi.org/10.1103/PhysRevC.92.034903}
  {\path{doi:10.1103/PhysRevC.92.034903}}.

\bibitem{ALICE:2016kpq}
J.~Adam, et~al., {Correlated event-by-event fluctuations of flow harmonics in
  Pb-Pb collisions at $\sqrt{s_{_{\rm NN}}}=2.76$ TeV}, Phys. Rev. Lett. 117
  (2016) 182301.
\newblock \href {http://arxiv.org/abs/1604.07663} {\path{arXiv:1604.07663}},
  \href {http://dx.doi.org/10.1103/PhysRevLett.117.182301}
  {\path{doi:10.1103/PhysRevLett.117.182301}}.

\bibitem{Giacalone:2016afq}
G.~Giacalone, L.~Yan, J.~Noronha-Hostler, J.-Y. Ollitrault, {Symmetric
  cumulants and event-plane correlations in Pb + Pb collisions}, Phys. Rev.
  C94~(1) (2016) 014906.
\newblock \href {http://arxiv.org/abs/1605.08303} {\path{arXiv:1605.08303}},
  \href {http://dx.doi.org/10.1103/PhysRevC.94.014906}
  {\path{doi:10.1103/PhysRevC.94.014906}}.

\bibitem{Zhu:2016puf}
X.~Zhu, Y.~Zhou, H.~Xu, H.~Song, {Correlations of flow harmonics in 2.76A TeV
  Pb--Pb collisions}, Phys. Rev. C95~(4) (2017) 044902.
\newblock \href {http://arxiv.org/abs/1608.05305} {\path{arXiv:1608.05305}},
  \href {http://dx.doi.org/10.1103/PhysRevC.95.044902}
  {\path{doi:10.1103/PhysRevC.95.044902}}.

\bibitem{Qian:2016pau}
J.~Qian, U.~Heinz, {Hydrodynamic flow amplitude correlations in event-by-event
  fluctuating heavy-ion collisions}, Phys. Rev. C94~(2) (2016) 024910.
\newblock \href {http://arxiv.org/abs/1607.01732} {\path{arXiv:1607.01732}},
  \href {http://dx.doi.org/10.1103/PhysRevC.94.024910}
  {\path{doi:10.1103/PhysRevC.94.024910}}.

\bibitem{Zhou:2015eya}
Y.~Zhou, K.~Xiao, Z.~Feng, F.~Liu, R.~Snellings, {Anisotropic distributions in
  a multiphase transport model}, Phys. Rev. C93~(3) (2016) 034909.
\newblock \href {http://arxiv.org/abs/1508.03306} {\path{arXiv:1508.03306}},
  \href {http://dx.doi.org/10.1103/PhysRevC.93.034909}
  {\path{doi:10.1103/PhysRevC.93.034909}}.

\bibitem{Acharya:2018ihu}
S.~Acharya, et~al., {Anisotropic flow in Xe-Xe collisions at
  $\mathbf{\sqrt{s_{\rm{NN}}} = 5.44}$ TeV}, Phys. Lett. B784 (2018) 82--95.
\newblock \href {http://arxiv.org/abs/1805.01832} {\path{arXiv:1805.01832}},
  \href {http://dx.doi.org/10.1016/j.physletb.2018.06.059}
  {\path{doi:10.1016/j.physletb.2018.06.059}}.

\bibitem{ATLAS-CONF-2018-011}
ATLAS Collaboration, \href{https://cds.cern.ch/record/2318870}{{Measurement of the azimuthal
  anisotropy of charged particle production in Xe+Xe collisions at
  $\sqrt{s_{\mathrm{NN}}}$=5.44~TeV with the ATLAS detector}}, Tech. Rep.
  ATLAS-CONF-2018-011, CERN, Geneva (May 2018).

\bibitem{CMS-PAS-HIN-18-001}
CMS Collaboration, \href{https://cds.cern.ch/record/2318319}{{Charged particle angular
  correlations in XeXe collision at $\sqrt{s_{NN}} = 5.44$ TeV}}, Tech. Rep.
  CMS-PAS-HIN-18-001, CERN, Geneva (2018).

\bibitem{Voloshin:2007pc}
S.~A. Voloshin, A.~M. Poskanzer, A.~Tang, G.~Wang, {Elliptic flow in the
  Gaussian model of eccentricity fluctuations}, Phys. Lett. B659 (2008)
  537--541.
\newblock \href {http://arxiv.org/abs/0708.0800} {\path{arXiv:0708.0800}},
  \href {http://dx.doi.org/10.1016/j.physletb.2007.11.043}
  {\path{doi:10.1016/j.physletb.2007.11.043}}.

\bibitem{Yan:2014afa}
L.~Yan, J.-Y. Ollitrault, A.~M. Poskanzer, {Eccentricity distributions in
  nucleus-nucleus collisions}, Phys. Rev. C90~(2) (2014) 024903.
\newblock \href {http://arxiv.org/abs/1405.6595} {\path{arXiv:1405.6595}},
  \href {http://dx.doi.org/10.1103/PhysRevC.90.024903}
  {\path{doi:10.1103/PhysRevC.90.024903}}.

\bibitem{Sirunyan:2017fts}
A.~M. Sirunyan, et~al., {Non-Gaussian elliptic-flow fluctuations in PbPb
  collisions at $\sqrt{s_{NN}} = 5.02$ TeV}\href
  {http://arxiv.org/abs/1711.05594} {\path{arXiv:1711.05594}}.

\bibitem{Acharya:2018lmh}
S.~Acharya, et~al., {Energy dependence and fluctuations of anisotropic flow in
  Pb-Pb collisions at $ \sqrt{s_{\mathrm{NN}}}=5.02 $ and 2.76 TeV}, JHEP 07
  (2018) 103.
\newblock \href {http://arxiv.org/abs/1804.02944} {\path{arXiv:1804.02944}},
  \href {http://dx.doi.org/10.1007/JHEP07(2018)103}
  {\path{doi:10.1007/JHEP07(2018)103}}.

\bibitem{Aaboud:2017tql}
M.~Aaboud, et~al., {Measurement of longitudinal flow decorrelations in Pb+Pb
  collisions at $\sqrt{s_{\rm NN}}=2.76$ and 5.02 TeV with the ATLAS detector},
  Eur. Phys. J. C78~(2) (2018) 142.
\newblock \href {http://arxiv.org/abs/1709.02301} {\path{arXiv:1709.02301}},
  \href {http://dx.doi.org/10.1140/epjc/s10052-018-5605-7}
  {\path{doi:10.1140/epjc/s10052-018-5605-7}}.
  
  
  
  \bibitem{Maowu:2018QM}
M.~Nie (for the STAR Collaboration), these proceedings.

\bibitem{Schukraft:2012ah}
J.~Schukraft, A.~Timmins, S.~A. Voloshin, {Ultra-relativistic nuclear
  collisions: event shape engineering}, Phys. Lett. B719 (2013) 394--398.
\newblock \href {http://arxiv.org/abs/1208.4563} {\path{arXiv:1208.4563}},
  \href {http://dx.doi.org/10.1016/j.physletb.2013.01.045}
  {\path{doi:10.1016/j.physletb.2013.01.045}}.

\bibitem{Bilandzic:2013kga}
A.~Bilandzic, C.~H. Christensen, K.~Gulbrandsen, A.~Hansen, Y.~Zhou, {Generic
  framework for anisotropic flow analyses with multiparticle azimuthal
  correlations}, Phys. Rev. C89~(6) (2014) 064904.
\newblock \href {http://arxiv.org/abs/1312.3572} {\path{arXiv:1312.3572}},
  \href {http://dx.doi.org/10.1103/PhysRevC.89.064904}
  {\path{doi:10.1103/PhysRevC.89.064904}}.

\bibitem{Acharya:2017gsw}
S.~Acharya, et~al., {Systematic studies of correlations between different order
  flow harmonics in Pb-Pb collisions at $\sqrt{s_{\rm NN}}$ = 2.76 TeV}, Phys.
  Rev. C97~(2) (2018) 024906.
\newblock \href {http://arxiv.org/abs/1709.01127} {\path{arXiv:1709.01127}},
  \href {http://dx.doi.org/10.1103/PhysRevC.97.024906}
  {\path{doi:10.1103/PhysRevC.97.024906}}.

\bibitem{Sirunyan:2017uyl}
A.~M. Sirunyan, et~al., {Observation of Correlated Azimuthal Anisotropy Fourier
  Harmonics in $pp$ and $p+Pb$ Collisions at the LHC}, Phys. Rev. Lett. 120~(9)
  (2018) 092301.
\newblock \href {http://arxiv.org/abs/1709.09189} {\path{arXiv:1709.09189}},
  \href {http://dx.doi.org/10.1103/PhysRevLett.120.092301}
  {\path{doi:10.1103/PhysRevLett.120.092301}}.

\bibitem{Gajdosova:2018pqo}
K.~Gajdosova (for the ALICE Collaboration), these proceedings, \newblock \href {http://arxiv.org/abs/1807.02998} {\path{arXiv:1807.02998}}.

\bibitem{Yan:2014afa} 
  L.~Yan, J.~Y.~Ollitrault and A.~M.~Poskanzer,
  Phys.\ Rev.\ C {\bf 90}, no. 2, 024903 (2014)
  doi:10.1103/PhysRevC.90.024903
  [arXiv:1405.6595 [nucl-th]].

\bibitem{Adams:2003zg}
J.~Adams, et~al., {Azimuthal anisotropy at RHIC: The First and fourth
  harmonics}, Phys. Rev. Lett. 92 (2004) 062301.
\newblock \href {http://arxiv.org/abs/nucl-ex/0310029}
  {\path{arXiv:nucl-ex/0310029}}, \href
  {http://dx.doi.org/10.1103/PhysRevLett.92.062301}
  {\path{doi:10.1103/PhysRevLett.92.062301}}.

\bibitem{Acharya:2017zfg}
S.~Acharya, et~al., {Linear and non-linear flow modes in Pb-Pb collisions at
  $\sqrt{s_{\rm NN}} =$ 2.76 TeV}, Phys. Lett. B773 (2017) 68--80.
\newblock \href {http://arxiv.org/abs/1705.04377} {\path{arXiv:1705.04377}},
  \href {http://dx.doi.org/10.1016/j.physletb.2017.07.060}
  {\path{doi:10.1016/j.physletb.2017.07.060}}.
  
\bibitem{Wu:2018cpc} 
  X.~Y.~Wu, L.~G.~Pang, G.~Y.~Qin and X.~N.~Wang,
  Phys.\ Rev.\ C {\bf 98}, no. 2, 024913 (2018)
  doi:10.1103/PhysRevC.98.024913
  [arXiv:1805.03762 [nucl-th]].
  
  
 \bibitem{Naghmeh:2018QM}
N.~Mohammadi (for the ALICE Collaboration), these proceedings. 

  
\bibitem{Pang:2016vdc} 
  L.~G.~Pang, K.~Zhou, N.~Su, H.~Petersen, H.~St\"ocker and X.~N.~Wang,
  Nature Commun.\  {\bf 9}, no. 1, 210 (2018)
  doi:10.1038/s41467-017-02726-3
  [arXiv:1612.04262 [hep-ph]].
  
\bibitem{Huang:2018fzn} 
  H.~Huang, B.~Xiao, H.~Xiong, Z.~Wu, Y.~Mu and H.~Song,
  arXiv:1801.03334 [nucl-th].
  
\bibitem{Zhao:2017yhj} 
  W.~Zhao, H.~j.~Xu and H.~Song,
  Eur.\ Phys.\ J.\ C {\bf 77}, no. 9, 645 (2017)
  doi:10.1140/epjc/s10052-017-5186-x
  
\bibitem{Niemi:2015voa} 
  H.~Niemi, K.~J.~Eskola, R.~Paatelainen and K.~Tuominen,
  Phys.\ Rev.\ C {\bf 93}, no. 1, 014912 (2016)
  doi:10.1103/PhysRevC.93.014912
  [arXiv:1511.04296 [hep-ph]].
  
\bibitem{McDonald:2016vlt} 
  S.~McDonald, C.~Shen, F.~Fillion-Gourdeau, S.~Jeon and C.~Gale,
  Phys.\ Rev.\ C {\bf 95}, no. 6, 064913 (2017)
  doi:10.1103/PhysRevC.95.064913
  [arXiv:1609.02958 [hep-ph]].

\end{thebibliography}







\end{document}